# Pore pressure development in small-strain undrained loading of soils according to a simple model

José Jorge Nader

Department of Structural and Geotechnical Engineering, Polytechnic School, University of São Paulo

05508-900, São Paulo, Brazil, e-mail: jjnader@usp.br, 55-11-3091-5522, 55-11-3091-5181 (fax)

**Abstract:** This note discusses a simple equation that may be useful in the study of pore pressures generated in the undrained loading of soils. The equation is employed to describe the pore pressure development accompanying a small stress increment in the undrained triaxial compression test and in the undrained simple shear test. The cases of incompressible and compressible pore fluids are examined.

**Keywords**: pore pressure; undrained loading; constitutive model.

**1. Introduction**

To start the discussion about pore pressures developed during an undrained loading, the soil skeleton is usually assumed to be linearly elastic [2, 3]. In spite of its simplicity, the linear elastic model is quite useful since it clearly shows that the relation between a given undrained increment of total stress and the produced pore pressure depends on the volumetric compressibility of the soil skeleton and of the pore fluid. However, on the other hand, it is unable to describe an important feature of soil behaviour, namely, that in real soils shear stresses may cause volume changes and so, in an undrained loading, shear stresses may generate pore pressures. In order to represent this phenomenon, we examine, in this note, a simple small-strain equation derived from the one proposed in [6].

**2. The equation**

We assume that the effective stress increment tensor $\Delta\overline{\boldsymbol{\sigma}}$ is related to the infinitesimal strain tensor $\boldsymbol{\varepsilon}$ by

$$\Delta\overline{\boldsymbol{\sigma}} = (K\varepsilon_v - \omega G \varepsilon_d)\mathbf{I} + 2G\boldsymbol{\varepsilon_d} \qquad (1)$$

where $\varepsilon_v$ is the volumetric strain (the first invariant of $\boldsymbol{\varepsilon}$), $\boldsymbol{\varepsilon_d}$ is the deviator strain tensor (the deviatoric part of $\boldsymbol{\varepsilon}$), $\varepsilon_d$ is the Euclidean norm of $\boldsymbol{\varepsilon_d}$ ($\varepsilon_d = \sqrt{-2I_2(\boldsymbol{\varepsilon_d})}$, $I_2(\boldsymbol{\varepsilon_d})$ being the second invariant of $\boldsymbol{\varepsilon_d}$), $\mathbf{I}$ is the identity tensor, *K* and *G* are, respectively, the bulk modulus and the shear modulus of the soil skeleton, and ω is a parameter controlling contractancy or dilatancy. In $\Delta\bar{\boldsymbol{\sigma}}$ and $\boldsymbol{\varepsilon}$ the soil mechanics sign convention is employed (compressive stresses and strains are positive etc). Eq. 1 is a particular instance of the equation proposed in [6] for materials having equal compression and expansion bulk moduli. Besides, in section 5 it will be shown that eq. 1 can be seen as a small-strain hypoplastic equation. Note in addition that, when ω is zero, eq. 1 reduces to the well-known linear elastic equation.

We will see now that ω has a clear physical meaning; it is responsible for the occurrence of volume changes accompanying a pure shear stress increment. In fact, for

$$[\Delta\bar{\boldsymbol{\sigma}}] = \begin{bmatrix} 0 & 0 & 0 \\ 0 & 0 & \Delta\tau \\ 0 & \Delta\tau & 0 \end{bmatrix} \tag{2}$$

(with $\Delta\tau > 0$), eq. 1 yields

$$[\boldsymbol{\varepsilon}] = \begin{bmatrix} \varepsilon & 0 & 0 \\ 0 & \varepsilon & \gamma/2 \\ 0 & \gamma/2 & \varepsilon \end{bmatrix}, \tag{3}$$

with $\gamma = \Delta\tau/G$ (as in linear elasticity) and $\varepsilon = \omega\Delta\tau\sqrt{2}/6K$ (unlike linear elasticity, in which $\varepsilon$ would be zero). The volumetric strain is, therefore, $\varepsilon_v = \omega\Delta\tau\sqrt{2}/2K$. If $\omega > 0$, then $\varepsilon_v > 0$ (contraction). If $\omega < 0$, then $\varepsilon_v < 0$ (expansion).

In the next two sections, in order to illustrate the model response, we will employ eq. 1 to compute pore pressures caused by a stress increment in an undrained axial compression test and in an undrained simple shear test. The cases of incompressible and compressible pore fluid will be examined.

### 3. Undrained axial compression test

In an undrained axial compression test, a positive total axial stress increment ($\Delta\sigma_a$) is applied while the total radial stress increment is zero. Hence, $\Delta\sigma_a = \Delta\bar{\sigma}_a + \Delta u$ and $\Delta\bar{\sigma}_r = -\Delta u$ ($\Delta\bar{\sigma}_a$ is the axial effective stress increment, $\Delta u$ is the pore pressure increment and $\Delta\bar{\sigma}_r$ is the radial effective stress increment). The corresponding strain matrix reads

$$[\varepsilon] = \begin{bmatrix} \varepsilon_r & 0 & 0 \\ 0 & \varepsilon_r & 0 \\ 0 & 0 & \varepsilon_a \end{bmatrix}, \qquad (4)$$

where $\varepsilon_r$ is the radial strain and $\varepsilon_a$ is the axial strain ($\varepsilon_r < 0$, $\varepsilon_a > 0$). Thus $\varepsilon_v = 2\varepsilon_r + \varepsilon_a$ and $\varepsilon_d = \sqrt{2/3}(\varepsilon_a - \varepsilon_r)$.

If the pore fluid is incompressible, $\varepsilon_v = 0$, and, from eq. 1, we obtain the following result for the ratio $\Delta u / \Delta \sigma_a$ (the so-called parameter $A$):

$$\frac{\Delta u}{\Delta \sigma_a} = \frac{1}{3} + \frac{\omega}{\sqrt{6}}. \qquad (5)$$

If the pore fluid is compressible (elastic), $\Delta u = K_f \varepsilon_v / n$, where $K_f$ is the fluid bulk modulus and $n$ is the soil porosity. From eq. 1 we get

$$\frac{\Delta u}{\Delta \sigma_a} = \left( \frac{1}{3} + \frac{\omega}{\sqrt{6}} \right)\left( 1 + \frac{nK}{K_f} \right)^{-1}. \qquad (6)$$

Clearly, for $\omega = 0$ the response is that of an elastic soil skeleton. If $\omega > -\sqrt{6}/3$, positive pore pressures are produced. The opposite happens for $\omega < -\sqrt{6}/3$. Therefore, a material that contracts in drained pure shear ($\omega > 0$) develops here positive pore pressures. As for materials that dilate in drained pure shear ($\omega < 0$), three separate cases should be considered: if $\omega < -\sqrt{6}/3$, $\Delta u < 0$; if $\omega = -\sqrt{6}/3$, $\Delta u = 0$; if $\omega$ is in the interval $-\sqrt{6}/3 < \omega < 0$, $\Delta u > 0$.

## 4. Undrained simple shear test

On horizontal planes a shear stress increment $\Delta \tau$ is applied while the vertical total stress increment is zero: therefore $\Delta u = -\Delta \overline{\sigma}_v$ ($\Delta \overline{\sigma}_v$ is the vertical effective stress increment). As horizontal strains are zero inside the simple shear test apparatus, the strain matrix reads

$$[\varepsilon] = \begin{bmatrix} 0 & 0 & 0 \\ 0 & 0 & \gamma/2 \\ 0 & \gamma/2 & \varepsilon \end{bmatrix}, \qquad (7)$$

where $\varepsilon$ is the vertical strain and $\gamma$ is the shear strain. So $\varepsilon_v = \varepsilon$ and $\varepsilon_d = \sqrt{2\varepsilon^2/3 + \gamma^2/2}$. We will now compute the relation $\Delta u / \Delta \tau$ as predicted by the model.

If the pore fluid is incompressible, eq. 1 yields

$$\frac{\Delta u}{\Delta \tau} = \frac{\omega}{\sqrt{2}}, \tag{8}$$

whereas, if it is compressible, a more complex expression results, involving not only $K$ but also $G$:

$$\frac{\Delta u}{\Delta \tau} = \frac{\omega}{\sqrt{2}} \left\{ \left[1 + \frac{n}{K_f}\left(K + \frac{4G}{3}\right)\right]^2 - \frac{2}{3}\left(\frac{n\omega G}{K_f}\right)^2 \right\}^{-1/2}. \tag{9}$$

The relation with volume changes accompanying a drained pure shear stress increment is direct. A material that contracts in drained pure shear develops here positive pore pressures. A material that dilates in drained pure shear develops here negative pore pressures.

**5. Relation with hypoplasticity. Small-strain hypoplastic equations.**

The general equation of hypoplasticity reads [1]:

$$\dot{\mathbf{T}} = \mathbf{h}(\mathbf{T}, \mathbf{D}) - \mathbf{TW} + \mathbf{WT}, \tag{10}$$

where $\mathbf{T}$ is the effective stress tensor, $\mathbf{D}$ is the stretching tensor, , $\mathbf{W}$ is the spin tensor, $\dot{\mathbf{T}}$ is the time derivative of $\mathbf{T}$. The function $\mathbf{h}$ is positively homogeneous of degree one in $\mathbf{D}$ and isotropic in relation to both $\mathbf{T}$ and $\mathbf{D}$. Here the continuum mechanics sign convention is employed (compressive stresses and strains are negative etc).

With the aim of studying the response of hypoplastic equations near the reference configuration (in a neighbourhood of $t = 0$), we follow the same steps taken in [8] with regard to hypoelasticity and write $\mathbf{T}(t) = \mathbf{T}_0 + \dot{\mathbf{T}}(0)t + o(t)$, as $t \to 0$. From eq. 10: $\dot{\mathbf{T}}(0) = \mathbf{h}(\mathbf{T}_0, \mathbf{D}_0) + \mathbf{W}_0\mathbf{T}_0 - \mathbf{T}_0\mathbf{W}_0$ (the subscript 0 refers to values calculated at $t = 0$). Taking into account that $\mathbf{D}_0 = \dot{\mathbf{E}}_0$ e $\mathbf{W}_0 = \dot{\mathbf{R}}_0^*$ (**E** is the infinitesimal strain tensor and $\mathbf{R}^*$ is the infinitesimal rotation tensor), we arrive at

$$\mathbf{T}(t) = \mathbf{T}_0 + \mathbf{h}(\mathbf{T}_0, \mathbf{E}(t)) + \mathbf{R}^*(t)\mathbf{T}_0 - \mathbf{T}_0\mathbf{R}^*(t) + o(t). \tag{11}$$

By neglecting the higher order terms indicated by $o(t)$, an approximate equation for the hypoplastic behaviour under small-strains is obtained (as in [5]). Now, we assume that the initial stress is spherical (therefore, $\mathbf{R}^*(t)\mathbf{T}_0 - \mathbf{T}_0\mathbf{R}^*(t) = \mathbf{0}$) and introduce $\Delta\mathbf{T}(t) = \mathbf{T}(t) - \mathbf{T}_0$ to produce $\Delta\mathbf{T} = \mathbf{h}(\mathbf{T}_0, \mathbf{E})$. For a given spherical $\mathbf{T}_0$, define $\mathbf{h}_0(\mathbf{E}) = \mathbf{h}(\mathbf{T}_0, \mathbf{E})$. Thus

$$\Delta\mathbf{T} = \mathbf{h}_0(\mathbf{E}) \tag{12}$$

In view of the properties of $\mathbf{h}$, it is clear that $\mathbf{h}_0$ is isotropic and positively homogeneous of degree one. Finally, to obtain eq. 1 choose $\mathbf{h}_0(\mathbf{E}) = (Ktr\mathbf{E} + \omega G\|\mathbf{E_d}\|)\mathbf{I} + 2G\mathbf{E_d}$ and introduce $\Delta\mathbf{T} = -\Delta\bar{\boldsymbol{\sigma}}$ and $\mathbf{E} = -\boldsymbol{\varepsilon}$.

## 6. Final remarks

We have seen that, with the help of a simple equation, as is eq. 1, one can get closer to real soil behaviour, at least from a qualitative point of view, than with linear elasticity, as far as the development of pore pressures in undrained loading is concerned. Outside the realm of small strains (and corresponding small stress increments), when it is necessary to describe non-linear pore pressure-strain curves, more complex constitutive models should be used (as done, for instance, in [7], [9] and [4]).